\documentclass[journal=nalefd,manuscript=letter]{achemso}
\usepackage[version=3]{mhchem} 
\usepackage[T1]{fontenc}       
\usepackage[pdftex]{xcolor}

\author{Delphine Coursault}
\affiliation[JFI]
{James Franck Institute, University of Chicago, IL 60637, USA}
\author{Nishant Sule}
\affiliation[JFI]
{James Franck Institute, University of Chicago, IL 60637, USA}
\author{John Parker}
\affiliation[JFI]
{James Franck Institute, University of Chicago, IL 60637, USA}
\alsoaffiliation[uoc]
{Department of Physics, University of Chicago, IL 60637, USA}
\author{Ying Bao}
\affiliation[JFI]
{James Franck Institute, University of Chicago, IL 60637, USA}
\alsoaffiliation[uoc]
{Department of Chemistry, University of Chicago, IL 60637, USA}
\alsoaffiliation[WWU]
{Department of Chemistry, Western Washington University, WA 98225, USA}
\author{Norbert F. Scherer}
\affiliation[JFI]
{James Franck Institute, University of Chicago, IL 60637, USA}
\alsoaffiliation[uoc]
{Department of Chemistry, University of Chicago, IL 60637, USA}
\email{nfschere@uchicago.edu}

\title[An \textsf{achemso} demo]
  {Dynamics of optically directed assembly and disassembly  of plasmonic nanoplatelet arrays}

\keywords{Optical trapping, electrodynamic coupling, plasmonics, optical matter, nanoparticle, multipole, nanoplatelet.} 

\begin{document}

%
%
%
%

\begin{abstract} Studies of nanoparticle-based optical matter have only considered spherical constituents. Yet nanoparticles with other shapes are expected to have different local electromagnetic field distributions and therefore interactions with neighbors in optical matter arrays. Therefore, one would expect their dynamics to be different as well.  We investigate directed-assembly of ordered arrays of plasmonic nanoplatelets in optical line traps demonstrating reconfigurability of the array by altering the phase gradient via holographic beam shaping. The weaker gradient forces on and resultant slower motion of the nanoplatelets as compared with plasmonic nanospheres allows precise study of their assembly and disassembly dynamics. Both temporal and spatial correlations are detected between particles separated by some hundreds of nanometers to several microns. Electrodynamics simulations reveal the presence of multipolar plasmon modes that induce short range (near-field) and longer range electrodynamic interactions. These interactions cause both the strong correlations and the non-uniform dynamics observed. Our findings demonstrate new opportunities to generate complex adressable optical matter by exploiting interference between mutipolar plamon modes and create novel active optical technology.
\end{abstract}

 
Optical matter, an assembly whose constituents are held together only by electromagnetic interactions,\cite{Burns1990} has richness and complexity that is only beginning to be explored and exploited. At its heart is the concept of optical binding \cite{Burns1989, Dholakia2010} of the constituent elements (e.g. dielectric colloids, plasmonic nanoparticles).\cite{Righini2007, Demergis2012, Yan2014a, Yan2015} Whether in an incident plane wave or a shaped electromagnetic field, each (trapped) entity restructures the field creating forces that can cause more particles to be trapped and localized with wavelength scale separations creating mesoscale assemblies. Controlled phase and amplitude shaping of optical fields, e.g. with spatial light modulators (SLMs), allow tailoring optical forces to shape optical matter structures and drive their formation. Optical phase gradients in particular opens tremendous opportunities for optical manipulation.\cite{Grier2003, Roichman2008, Figliozzi2017} Therefore, optical manipulation of colloids is a growing field of research.\cite{Dienerowitz2008, Dholakia2010, Jones2015a}

While optical binding between two Rayleigh particles was reported in 1994,\cite{Dapasse1994} it has only recently been harnessed  to direct and drive nanoparticle assembly. \cite{Marago2013}  In particular, 1D, 2D and 3D assemblies of spherical plasmonic nanoparticles (as small as 40 nm) into optically-bound arrays and supra-crystals have been achieved.\cite{Yan2013, Yan2014a, Yan2015} The localized surface plasmon resonance properties of Ag or Au nanoparticles increase their scattering cross-sections thus enhancing electromagnetic interactions and propensity for nanoscale manipulation.\cite{Demergis2012,Yan2013, Yan2014} 

Optical binding interactions are very sensitive to light polarization,\cite{Yan2014a} which has created opportunities for controlling the dynamics in driven optical matter.\cite{Figliozzi2017}. It has been shown that polarization of the electromagnetic field induces anisotropic forces and torques on anisotropic particles,\cite{Messina2015, Brzobohaty2015, Tong2010, Yan2013b, Yan2013b, Liaw2015}. Thus it is expected that anisotropic nanoparticles should allow creating new optical matter structures with novel dynamics due to anisotropic interactions. However no experimental studies of optical binding or formation of optical matter from anisotropic (nano-)particles have been reported. While near-field interactions have been investigated for small plasmonic (gold) nanowires ($d/\lambda$<<1),\cite{Zhao2010, Ekeroth2016} optical binding has only been investigated theoretically for large dielectric nanorods, where ladder-like structures with edge-to-edge optical binding separations have been predicted for a range of nanorod aspect ratios.\cite{Simpson2017} We expect new properties to emerge in optical matter constituted of anisotropic nanoparticles due to the increased significance of higher order (scattering) modes and potentially different or enhanced many-body effects. 

In this letter, we demonstrate that highly anisotropic nano-objects, gold nanoplatelets (Au-NPLs) with an aspect ratio $\geq$10, can form robust optical matter structures, but with very different interactions and dynamics than nanospheres as manifest  both in the steady-state dynamics of the NPL arrays and during the assembly/disassembly process. We use linearly polarized optical line traps\cite{Yan2015} to investigate optical assembly of Au-NPLs in a quasi-1D geometry. The use of optical line traps allows direct comparison of the influence of the phase gradient on NPLs and to those previously reported for to plasmonic nanospheres. \cite{Yan2015} To further manipulate  and stabilize NPL interactions, we superimposed a Gaussian beam, the so-called zero-order, to the center of the line trap as manifest in NPL separations and fluctuations. 

\begin{figure}[H]
\includegraphics[width=1\columnwidth]{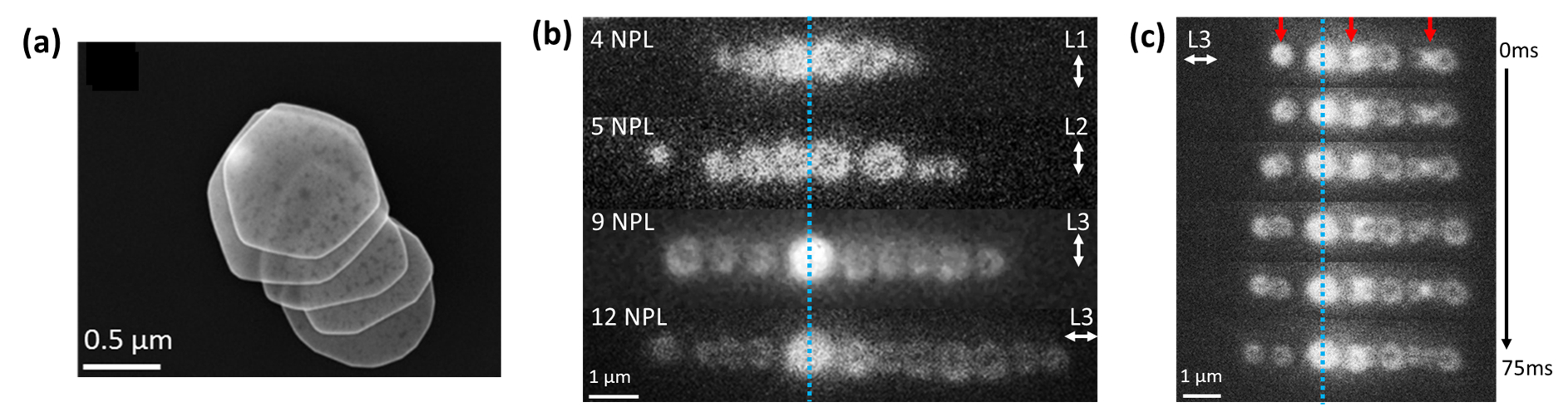}
\caption{Shapes and sizes of Au NPLs and arrays in optical line traps. a) TEM image of Au NPLs. b) and c) Dark field images of Au NPL linear arrays in the line trap. The white double arrows indicate the direction of light polarization in the optical traps. The blue dotted line indicates the position of the zero order. In (b) from L1 to L3, we observe formation of linear arrays with the number of NPLs increasing with the length of the line trap. In (c) the red arrows point at the areas where NPLs overlap; i.e., they stack on top of each other but do not stick to each other. The six images are successive frames from the video.}
\label{fig:chain2}
\end{figure}

In our optical trapping setup (schematic in Supporting Information), the output from a cw Ti:Sapphire laser, 70 mW power measured before the objective, is directed to an inverted microscope (Olympus) and focused into a coverslip sandwich sample cell with a high numerical aperture water immersion objective (NA=1.2). The radiation pressure exerted by the laser beam pushes the particles towards the top coverslip of the sample cell where they are trapped near the water/glass interface. We use a spatial light modulator (SLM; Hamamatsu) to shape the Gaussian beam into a line trap by applying phase masks that act as cylindrical lenses that can be either concave (type I) or convex (type II) (see Supporting information, Figure S1b). The beam's intensity distribution remains Gaussian along the trap (Supporting information, Figure S1c), while the phase exhibits a parabolic distribution either positive (type I) or negative (type II) (Supporting information, Figure S1d).  Consequently, the optical gradient force can be decomposed into two forces: (i) the intensity gradient force, which is the same  for the two types of traps, while (ii) the phase gradient force is of opposite sign for type I and II optical traps.\cite{Grier2003, Roichman2008} The balance between the forces depends on the particle's dielectric properties. If the phase gradient dominates over the intensity gradient, particles can be driven directionally either into or out of the trap when switching the sign of the phase gradient.\cite{Yan2015}.
 When reflecting from the SLM, 5$\%$ of the incident Gaussian beam is not phase modulated and becomes the so-called ``zero-order" beam. The resulting line trap and zero-order beams are shown in the Supporting Information while the consequences on NPL trapping are shown in Figure~\ref{fig:chain2}b. In some experiments, we superimposed the zero-order focused Gaussian beam onto the optical line trap to enhance the electromagnetic field at the center of the trap (optical images shown in Supporting Information). 


\textbf{Formation of nanoplatelet linear arrays}. Figure~\ref{fig:chain2}a shows a transmission electron microscopy (TEM) image of Au-NPLs. The shape and diameter of NPLs vary, but on average they are disk-like colloids with a diameter of 700nm and a 25nm thickness. 

\begin{figure}[H]
\includegraphics[width=1\columnwidth]{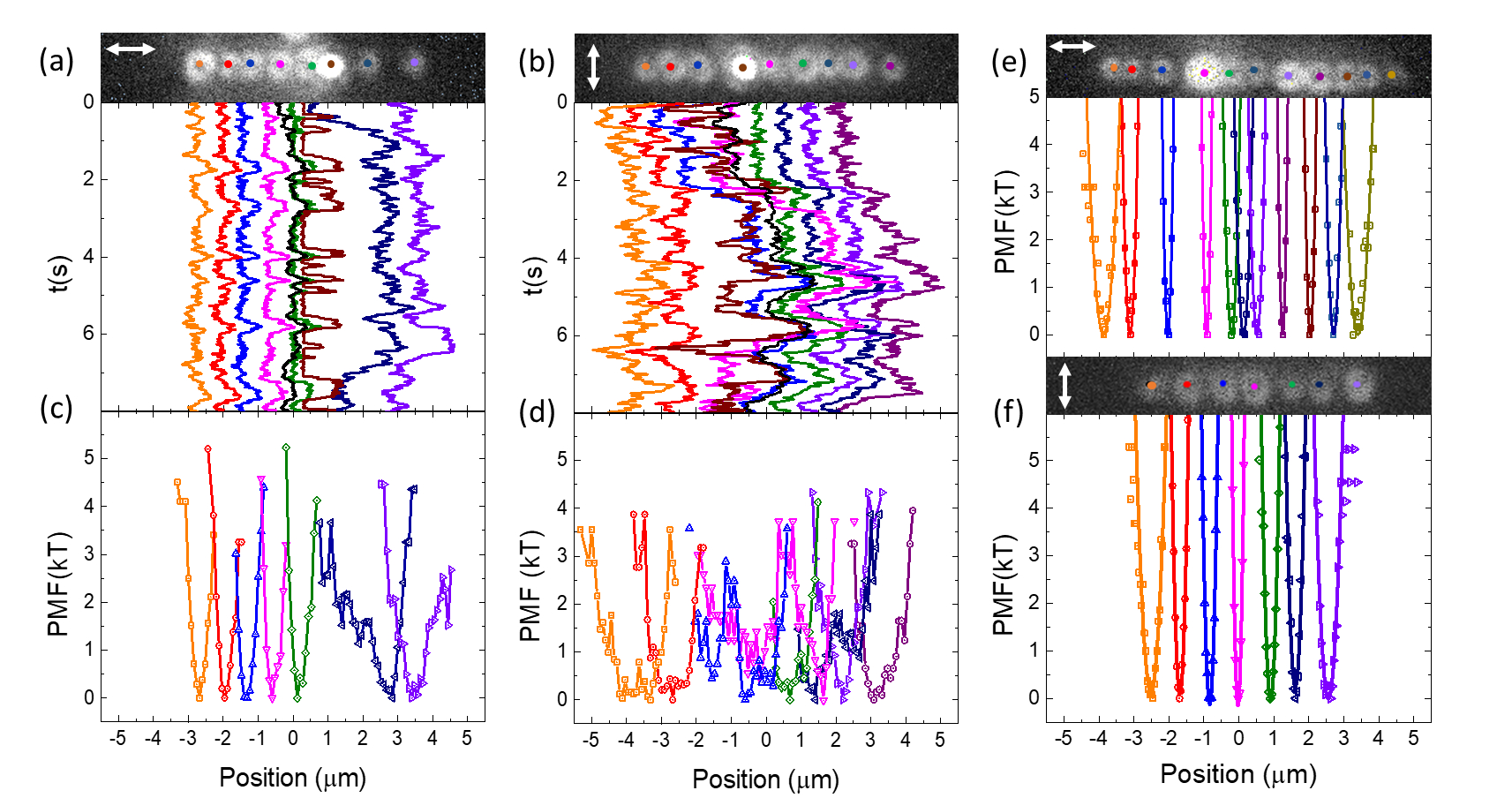}
\caption{Dark field images and time trajectories of the NPLs along the line trap in the absence of the zero-order beam for polarisations (a) parallel (b) perpendicular to the NPL array. The black and brown trajectories represent the mean of the NPLs and that of a single Au nanosphere, respectively. (c) and (d) are the potentials of mean force (PMFs) calculated from the probability density distributions of the time trajectories in absence of the zero-order beam. (e) and (f) are the PMFs of NPL arrays in the presence of the zero-order beam for both polarizations. (Data are shown in the Supporting Information). NPL 4 (pink) is located at the position of the zero-order beam for both polarizations. The corresponding trajectories are shown in the Supporting Information. Symbols are experimental data and solid lines are the corresponding  harmonic fits (only shown in the presence of the zero-order beam; i.e., only in (e) and (f)).}
\label{fig:pmf}
\end{figure}

\begin{figure}[H]
\includegraphics[width=1\columnwidth]{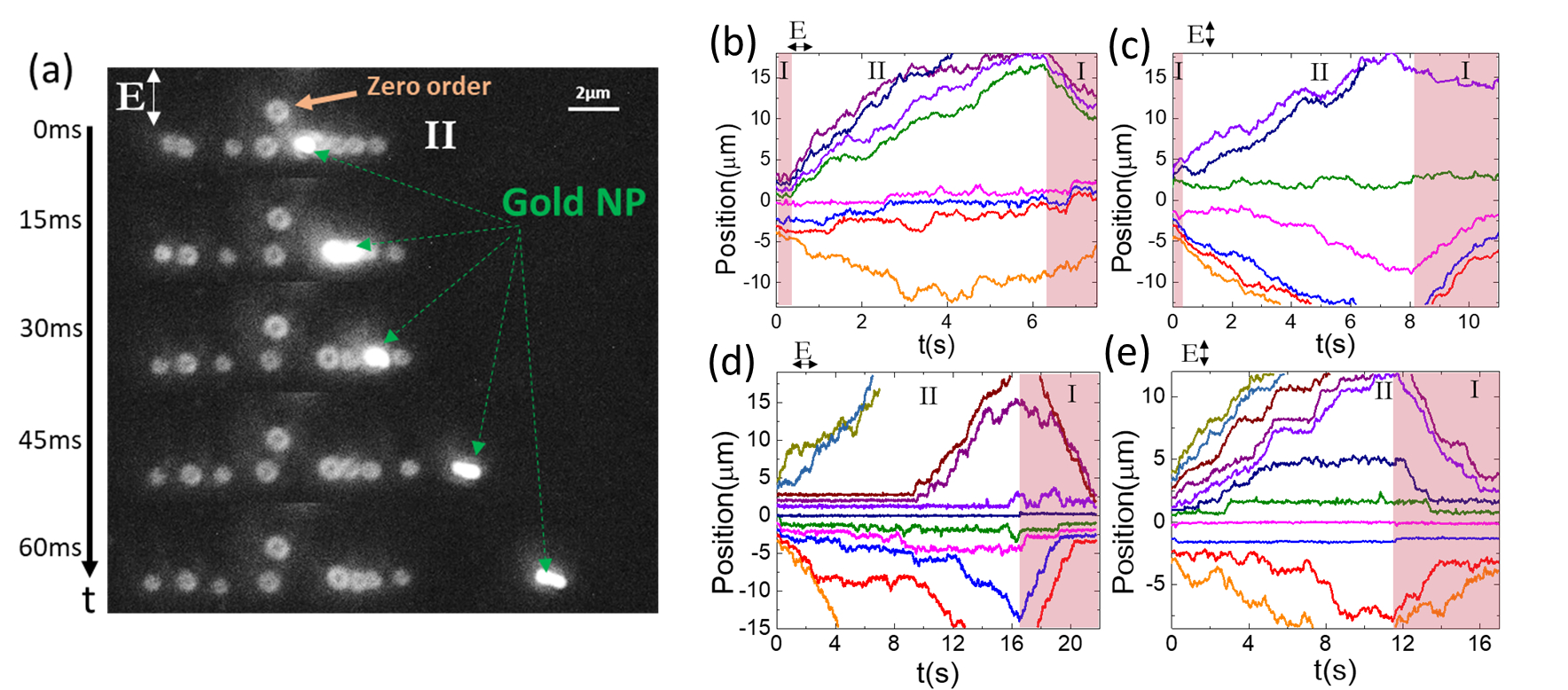}
\caption{Driven dynamics of NPLs in line traps a) Successive dark field images of NPLs and a single nanosphere driven out of the trap. Time trajectories of NPLs driven along the main axis in type II and then type I trap (shaded regions) respectively for perpendicular and parallel polarization direction in absence (b-c) and presence (d-e) of the zero-order beam. Note that the shape of the Au nanosphere is elongated due to its rapid motion during the 15 ms acquisition time per frame.}
\label{fig:dyn1}
\end{figure}

Viewed by dark field microscopy (Figure~\ref{fig:chain2}b-c), NPLs exhibit a donut-like shape with a bright corona around a dark central spot and appear with a weak scattering intensity (weak as compared with Au or Ag nanosphere as shown in Figure~\ref{fig:dyn1}a-b) for the spectral detection window of the dark-field microscopy (450-750nm). The NPLs are not only confined in the line trap but they are oriented perpendicular to the beam propagation direction (parallel to the  cover slip). They organize into linear array for different optical trap configurations: (i) with linearly polarized light either parallel or perpendicular to the line trap axis, (ii) when the line trap and the Gaussian trap (zero-order) are superimposed (Figure~\ref{fig:chain2}b-c) or shifted from each other (Figure~\ref{fig:pmf}). They remain in the linear configuration when the optical phase gradient is either positive (type I trap) or negative (type II trap) (see Figure~\ref{fig:dyn1}). 
Figure~\ref{fig:chain2}b shows that NPLs can form stable linear arrays in a type I trap (i.e. for an inward directed phase gradient), with array lengths increasing from  4 NPLs up to 12 NPLs. The NPLs are very close to one another with a center-to-center distance with their nearest neighbor being essentially equal to their diameter (d=700nm) on average. As illustrated in Figure~\ref{fig:chain2}c, we observe small variations of their scattered intensity  that we attribute to two different phenomena: (i) fluctuations in their orientation, and (ii) partial or full lateral overlap. Generally, the NPLs exhibit very small fluctuations of their orientation from perpendicularity to the incident beam's wave-vector. However, when a  larger variation of their orientation occurs, which can be induced by making the incident beam circularly polarized, a more drastic change in the brightness and contrast occurs (see Supporting Information). 

We also observe that the NPLs can overlap laterally. NPLs can slide on top of one another and can even remain stably superimposed. The region of partial overlap appear brighter in the dark field images as shown in Figure~\ref{fig:chain2}b-c. In particular, the NPLs tend to remain stably fully overlapped when trapped at the zero-order location.  This overlap phenomenon is observed for both linear polarizations, parallel and perpendicular to the line trap axis. Also, as shown in Figure~\ref{fig:chain2}c, we observe  total or partial overlapping of NPLs with one or two neighboring NPL  when the polarization is parallel to the trap direction. The red arrows in Figure~\ref{fig:chain2}c point to these overlaps. Partial overlap is less likely to occur when the polarization direction is perpendicular to the trap axis. It occurs more frequently when  a NPL is already trapped by the zero-order beam and a newly trapped NPL is inserted within the array.

From these observations, the interactions between NPLs appear to be  different from what has been reported either for large colloids ($d/\lambda>1$) and  even 200 nm Au nanospheres that are closer to the Rayleigh criteria ($d/\lambda\approx$ 0.8). Strong optical binding has been demonstrated for both micron-sized and submicron-sized particles with inter-particle distances equal to multiples of the wavelength of the incident beam in the refractive index of the host medium.\cite{Burns1989, Dholakia2010, Demergis2012} Smaller plasmonic nanoparticles with a diameter ranging from 40 nm to 200 nm can be organized on mesoscales spaced at the wavelength of light by optical binding. Electrodynamically bound dimers of spherical nanoparticles can form through near-field interactions aligned parallel to the light polarization.\cite{Yan2014a, Sule2017} Such point dipole-like interactions do not occur in the case of NPLs ($d/\lambda>1$).
 
\begin{figure}[H]
\includegraphics[width=1\columnwidth]{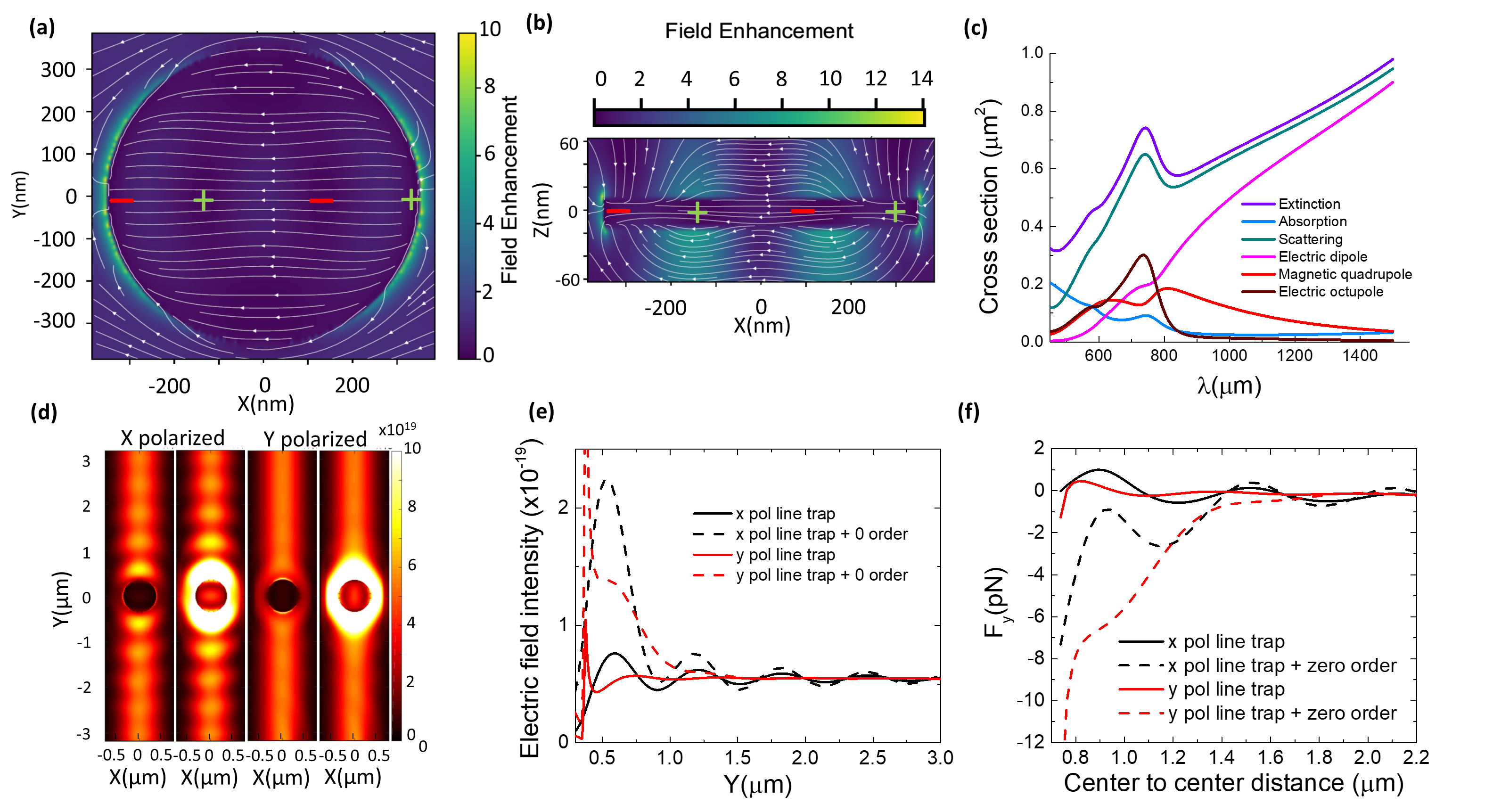}
\caption{FDTD simulations of a Au NPL (disk of diameter 700 nm and height  of 25 nm) in n=1.33 for water: (a,b) Map of the electric field enhancement at $\lambda$=800nm (vacuum) for incident light propagating along the z direction and polarized along the x axis (a) in the XY plane, parallel to the plate (b) in the XZ plane perpendicular to the NPL. (c) Extinction, absorption, and scattering cross sections with modal decomposition of the scattering cross section. (d) 2D intensity map of the line trap with and without the presence of the zero-order beam with one NPL trap at the center of the beam for a polarization direction parallel (x-polarized) or perpendicular to the line trap (y-polarized). (e) Electric field intensity along the line trap when a NPL is trapped at the center of the beam. (f) Comparison of the optical force along the y-axis when a NPL is trapped at the center of the line trap in the absence and presence of the zero-order beam.}
\label{fig:FDTD}
\end{figure}

\textbf{Influence of near field interactions}. We performed electrodynamics (finite difference time domain, FDTD) simulations to help understand why NPLs behave similaly and differently than nanospheres. Figure~\ref{fig:FDTD}a shows the calculated total field distribution in the NPL plane (XY plane-z=0). At first glance, the field  resembles  a dipole with strong enhancement at the edge of the NPL parallel to the light polarization; an enhancement 2 to 9 times stronger than perpendicular to the polarization. A smaller quadrupole-like enhancement pattern is also visible. This observation is evident in the XZ plane (Figure~\ref{fig:FDTD}b): the field enhancement shows two nodes and anti-nodes and the phase changes sign 4 times over the NPL, which is indicative of retardation. Figure~\ref{fig:FDTD}c shows the extinction cross-section of a NPL that is dominated by the scattering contribution. Three multipolar plasmon modes of NPLs modes are identified by FDTD simulations using a spherical (vs cubical) flux box projecting onto multipolar basis functions.\cite{Parker2017} The three main modes in the light scattered by the NPL are: a spectrally very broad electric dipole, a magnetic quadrupole and an electric octupole. All 3 modes are excited at $\lambda$=800nm (laser excitation wavelength), but the electric dipole and magnetic quadrupole are dominant over the octupole mode. The electric field distribution and multipolar excitations suggest that strong near-field interactions occur between NPLs explaining why significant spatial overlapping of NPLs can occur, particularly when the polarization is set parallel to the line trap direction. 
\textbf{Influence of the zero-order focused beam}. We also performed FDTD simulations to shed light on the NPL overlap phenomena at the location of the zero-order beam. The results are presented in Figure~\ref{fig:FDTD}d-f. First, the optical line trap is simulated for both polarizations in the absence and presence of the zero-order beam with one NPL located at its center (Figure~\ref{fig:FDTD}d-e). The light scattered by the NPL strongly interferes with the incident beam for a light polarization direction perpendicular to the line the trap, while the interference effect is very weak for the parallel polarization direction. The electromagnetic field is strongly enhanced around and on top of the NPL in the presence of the zero-order beam for both polarizations. The amplitude of the first 3 interference nodes  is enhanced in the perpendicular polarization case. For parallel polarization, the enhancement decays to 0 on a similar length scale, around 2.4 $\mu$m away from the NPL center (4 times the wavelength in water). As illustrated in Figure~\ref{fig:FDTD}f, this strong field enhancement give rise to an enhanced optical gradient resulting in a strong attractive force toward the center. Then NPL interactions and inward-directed forces contribute to and cause the NPL stacking at the center of the trap in presence of the zero-order beam. Conversely, only partial overlapping was observed in absence of the zero-order beam (i.e., one plate overlapped with two other NPLs at the center of the trap).

\textbf{Correlated motions of NPLs in linear arrays at steady state}. The optical field-induced stability of nanoplatelet arrays result from the NPL/optical trap interactions and the intra-array interactions (interactions between NPLs). The relative significance of each can be established by comparing the behavior of NPLs within the array at steady-state, in absence and presence of the zero-order beam, for both polarization parallel and perpendicular to the line trap.
We tracked the motions of the NPLs for the four trap configurations; i.e., both polarizations and with and without the zero-order beam. In Figure~\ref{fig:pmf}, each trapped NPL is represented by a different color (colors are related to the position of the particle but do not imply identity of the NPL between the four experiments). The distances between the NPLs are comparable to their diameter for both polarizations (Table~\ref{tbl:dis}, Figure~\ref{fig:pmf}), and are not strongly seem to be affected  by the presence of the zero-order beam. On average, the inter-NPL distances are slighly smaller when the light polarisation is parallel to the trap, in agreement with a stronger near-field interaction parallel to the polarisation. The spacings, particularly for NPLs near the location of the zero-order beam, are in part the result of the locations of the maxima in the field intensity. The FDTD simulation results shown in Figure~\ref{fig:FDTD}f indicate that the  first $|E^{2}|$ maximum occurs at 1.1$\mu$m (1.7$\mu$m) in absence (presence) of the zero-order beam. The other spacing are very close edge-to-edge suggesting that significant even dominant interactions between the NPLs are near-field interactions. 
\begin{table}[H]
  \caption{Mean distance between the NPL in the absence (presence) of the zero-order beam. A,B are without and C,D are with the zero-order beam.*~indicates NPL pairs closest to the position of the zero order. }
  \label{tbl:dis}
		\begin{tabular}{c}
		\hline
		Distances between NPL($\mu$m)\\
				\begin{tabular}{llllllllllllll}
		Polar. & &1-2 &2-3 &3-4 &4-5 &5-6 &6-7 &7-8 &8-9 &9-10 &10-11\\
		\hline
		// &A &0.74 &0.64 &0.73 &0.73 & &0.57 & & & &\\
		$\bot$ &B &0.84 &0.87 &1.13 &1.1 &0.75 &0.77 & &0.89 & &\\
		// &C &0.77 &1.05 &1.1* &0.74* &0.32 &0.42 &0.71 &0.77 &0.66 &0.75\\
		$\bot$ &D &0.81 &0.87 &0.81* &0.92* &0.72 &0.92 & & & & &\\
		\hline
  \end{tabular}
		\end{tabular}
	\end{table}

Figures \ref{fig:pmf}a-b show the tracked positions of an 8 NPL array and a 7 NPL array, formed in absence of the zero-order beam, respectively, for polarizations perpendicular and parallel to the trap. A Au nanosphere (a side product of the synthesis) is also trapped near the center of the trap. Surprisingly, it can overlap with the NPLs but not get stuck to the NPLs \cite{Figliozzi2017}. The well defined (harmonic) PMFs indicate constrained NPL motion observed in the trajectories of Figure \ref{fig:pmf} about ``lattice sites''. However, the spacings are  not consistent with traditionnal optical binding\cite{Burns1990,Demergis2012,Yan2014a, Sule2015} in that the values shown in Table~\ref{tbl:dis} do not equal integral multiples of the wavelength of the optical trapping beam in water (n=1.33).\cite{Yan2013}

The trajectories also reveal strongly correlated motions of the NPLs suggesting that NPL arrays behave like rigid bodies. This observation is in qualitative agreement with prior results obtained for small spherical nanoparticles.\cite{Yan2015} In optical matter, oscillations of constituents about their mean locations have been described by analogy with phonons in a crystal lattice\cite{Abajo2007}: the lattice periodicity is dictated by the periodicity of the optical field while oscillations are induced by small perturbations that lead to small displacements of the particles from their lattice site. At equilibrium (or steady state) the spectrum of fluctuations could be associated with a lattice temperature.\cite{Figliozzi2017} The displacements are transmitted from one particle to another via their electrodynamic interactions, but there could also be a hydrodynamic interaction via the fluid (water).

In our experiment, we expect the intrinsec properties of both the trap (the optical intensity and phase gradients) and the NPLs (multipolar scattering), to affect any collective oscillations (modes) of the array. Because the trapping strength increases toward the trap center (see Supporting Information) and the number of neighbors is different for each NPL due to the finite size of the system, the amplitude of the fluctuations around the mean position of each NPL should change with the particles' distance from the center of the trap. Also, the shape of the NPLs will affect how they scatter light and therefore the nature of their electrodynamic interactions and thus the transmission of a perturbation from site to site including retardation. However, the maim point is the extensive correlation of NPL motion observed in the trajectories of Figure \ref{fig:pmf}.

Finally, we note that the presence of the Au nanosphere disrupts the NPL array. It clearly induces an asymmetry in the array and causes large amplitude fluctuations in the NPL trajectories. Indeed, as shown Figure~\ref{fig:pmf}, each large jump observed in the sphere trajectory is followed by large amplitude fluctuations in the NPL trajectories. The closer the NPLs are to the nanosphere the stronger the perturbation (i.e., the larger the amplitude of NPL displacement).
Interestingly, the response of the NPLs to each perturbation is delayed by several tens of milliseconds, highlighting the different scales of forces associated with the two types of nano-objects.
 
\textbf{Influence of zero-order beam on stabilizing optical matter arrays}. Given all the aformentionned observations, the persistence of  collective motions must imply that the interaction between the nanoplatelets is strong compared to k$_{B}$T and long range. The trajectories (see Supporting Information) show that the NPLs are more tightly trapped in the presence of the zero-order beam, which is particularly striking  near the center of the trap. This is in agreement with FDTD simulations (Figure~\ref{fig:FDTD}e), which show that interferences between the light scattered by the NPL and the optical trap are reinforced with the presence of the zero-order beam. Therefore, the collective oscillations observed in Figure~\ref{fig:pmf} (and the Supporting Information) are of smaller amplitude since these NPLs are more tightly bound to their ``sites''. Despite some asymmetry of the assembly, likely caused by the strong anchoring of one NPL at the zero-order location (whose position is shown by the pink color in Figure~\ref{fig:pmf}e-f), the arrays as a whole are more stabilized by the presence of the zero-order beam.

The influences of the zero-order beam and the polarization on the stability and fluctuations of the NPL arrays are quantified by the potential of mean force (PMF) of each individual NPL (Figure \ref{fig:pmf}) and of the mean trajectory of each array (see Supporting Information). Since the system of NPLs is at steady state (so long as no particles enter or leave the trap, we can determine the potential of mean force (PMF) confining the particle (NPL) from its probability density distribution, $PMF(\left\{x\right\}_{i})=-ln\left[P(\left\{x\right\}_{i})\right]$, where $\left\{x\right\}_{i}$ are the set of $x$ coordinate positions of particle i and $P(\left\{x\right\}_{i})$ is the associated probability density distribution.

\begin{table}
  \caption{Force constants of NPLs from harmonic fits to PMFs of Figure~\ref{fig:pmf}. A,B are without and C,D are with the zero-order beam.}
  \label{tbl:spring}
	\begin{tabular}{c}
	\hline
		k (pN/$\mu$m)\\
				\begin{tabular}{llllllllllllll}
		Polar. & &mean  &1 &2 &3 &4 &5 &6 &7 &8 &9 &10 &11 \\
				\hline
		// &A &0.04 &0.07 &0.09 &0.09 &0.01 &0.1 &\textbf{0.01} &\textbf{0.02} & & &\\
		$\bot$ &B &\textbf{0.01} &0.01 &0.02 &0.01 &$\oslash$ &0.02 &$\oslash$ &0.02 &0.02 & & &\\
		// &C &0.23 &0.08 &0.10 &0.69 &0.87 &0.33 &0.28 &0.32  &0.91 &0.64 &0.31 &0.09\\
		$\bot$ &D &0.68 &0.12 &0.38 &0.45 &0.76 &0.34 &0.27 &0.11 & & & &\\
				\hline
  \end{tabular}
	\end{tabular}
\end{table}

The trap stiffness for each individual NPL in each array is determined by fitting their associated PMF with a harmonic potential.\cite{Jones2015a}. The stiffness at their sites along the trap vary between approximately 10 fN/$\mu$m to 100 fN/$\mu$m (see Table~\ref{tbl:spring}), while the confinement perpendicular to the trap (not shown) is about 10-20 fN/$\mu$m. This might be viewed as surprising since the increase of the total power due to the zero order is only 5\%. We attribute the increase of the stiffness by a factor of 10 for both polarizations in the presence of the zero-order beam to the associated enhanced fields and inter-NPL interactions (see Figure \ref{fig:FDTD}. The influence of the number of NPLs in the trap on the stiffness of the array is also a factor.     

Both Figure~\ref{fig:pmf}e-f and Table~\ref{tbl:spring} show that the NPL-associated PMFs become wider (i.e. smaller stiffness) as the distance to the mean position (= 0 $\mu$m) increases (see Supporting Information) in line with a decreasing number or total absence of neighbors on their outward side. This observation is in agreement with a decrease of the optical intensity and optical gradient force along the line\cite{Yan2015}; both are enhanced by the presence of the zero-order beam (Figure~\ref{fig:FDTD}d). Their enhancements lead to stronger inter-NPL interactions that stabilize the overall array and each NPL in it. This conclusion is most dramatic when one NPL is strongly anchored at the zero order location (in pink Figure~\ref{fig:pmf}e-f). In one case, the center of the trap overlaps with the center of the array, (Figure \ref{fig:pmf}f). In the other case, because of the asymmetric assembly, the zero-order  beam did not overlap of the array center. It seems that the NPL at the zero-order position acts like a barrier that disconnects the motion of the NPL on each side (see Supporting Information). The induced electrodynamic asymmetry will require a deeper investigation, and is beyond the scope of this paper.  

Finally, we note that when the NPL array was disrupted by the presence of a Au nanosphere, the probability density distribution of the closest NPL departed from the expected Gaussian shape becoming either a strongly skewed Gaussian distribution (Figure \ref{fig:pmf}c) that leads to an asymmetric PMF, or a very broad distribution (Figure~\ref{fig:pmf}d) leading to multiple and shallow wells in the PMF. In these cases, either no value or only a rough estimation of the stiffness (bold) is given in Table~\ref{tbl:spring}.
 
\textbf{Influence of the phase gradient on NPL dynamics}. 
 We inverted the phase gradient (type II trap) after NPLs are organized into 1-D arrays, then waited 6-16 seconds and reversed the phase gradient again to a type I trap to measure their collective dynamics during dissassembly and assembly. This procedure, as shown in Figure~\ref{fig:dyn1}, drives the NPLs outward (type II trap) or inward (type I trap) along the main axis of the line trap. The influence of the zero-order beam is revealed when it is shifted away from the line trap. We study the dynamics for parallel and perpendicular polarizations. Figure~\ref{fig:dyn1}a shows consecutive snapshots (images) of the NPLs in a type II trap in which the phase gradient is opposite to the intensity gradient; that is, the phase gradient pushes the Au nanosphere and the Au NPLs out of the line trap along its main axis. The green arrows point at the single rapidly moving Au nanosphere that had been trapped together with the Au NPLs. The Au nanosphere is driven more than 10 times faster than the outer most nanoplatelet. The much slower motion of NPLs shows that they have much less responsive to the phase gradient than the Au nanospheres. 
We calculated the translational drag coefficient for a NPL oriented with its short axis perpendicular to the motion to determine whether hydrodynamic drag forces may be responsible for the slower motion of the NPLs. According to Perrin's formula \cite{Perrin1934}, the friction coefficient for an oblate ellipsoid is:
\begin{eqnarray}
	\gamma=32\pi\eta\frac{a^2-b^2}{(2a^2-3b^2)S-2a}\text{ with $a<b(=c)$, }\\
  \text{ where }S=\frac{2}{\sqrt{b^2-a^2}}\arctan\frac{\sqrt{b^2-a^2}}{a}
	\label{eq:perrin}
\end{eqnarray}
For a disk of height a=12.5~nm and a radius b=350~nm immersed in water ($\eta_{H_{2}O}$=$10^{-3}$~Pa.s) $\gamma_{NPL} =0.66\times10^{-12}$~(N.s)/m vs $\gamma_{NP}=2\times10^{-9}$~(N.s)/m for a 200nm diameter nanosphere. Thus, assuming equal electrodynamic forces, the hydrodynamics suggest that a nanosphere would move slower than a NPL (by 10$^{3}$ times). Even a correction that takes into account the distance to the substrate would not decrease the sphere drag by 3 orders of magnitude. Therefore, we conclude that the different behaviours come from the electrodynamic forces. Confirming insights from our FDTD simulations that the NPLs interact very differently with light than nanospheres.

Owing to the slower drift of the NPLs, the traps are rapidly cleared of incidental Au nanospheres thus allowing unperturbed measurements of NPL interactions and dynamics. The trajectories of the NPLs shown in Figure~\ref{fig:dyn1}b-c reveal non-monotonic and time-correlated drift. Despite the differences in the trajectories of the individual NPLs, there are several general or common features. First, the closer the NPLs were to the center, the  longer it took them to escape. Second, we observed that in most cases the NPLs did not escape and remained close to the center during the whole experiment. Third, when the NPL were driven outwards, they escape the trap (type II) at slower speed than when they are driven toward the center (type I trap). The trend is consistent with the fact that  intensity and phase gradients are the same (type I) or opposite (type II) sign. Now to the particulars. In the case of only a single trapped NPL (see Supporting Information), we observed that once the NPL escaped the trap center (type II), that it moves at constant speed, similar to the case of single nanospheres but slower\cite{Yan2015}. By contrast, when multiple NPLs are trapped, some portions of the single NPL trajectories appear highly correlated with neighbors; e.g. several NPLs move in near lock-step in panels (b)-(e) and/or exhibit halting plateau like correlated motions as in (e). The drift of each NPL appears altered according to the number of neighboring NPLs, the distance between them and their position in the trap. The correlated drift is most likely due to electrodynamic interactions between NPLs and with the spatially structured field in optical trap (see Figure \ref{fig:FDTD}d). We expect the hydrodynamic interactions between NPLs to play a minor role, particularly as it was recently shown that electrodynamic interactions can be dominant over hydrodynamic in light driven nanoparticle system.\cite{Figliozzi2017}.  

It is very surprising to observe correlation between them when they are more than 10$\mu$m away from the trap center. The effect is even more striking in the presence of the zero-order beam (Figure \ref{fig:dyn1}d-e); not only there are time correlations between the trajectories, but also spatial correlations with the NPLs pausing at regular distances from the center of trap when they are driven. We believe that the interferences between incident and scattered light sculpt the potential landscape \textit{in situ} over tens of micrometers offering new opportunity for tuning particle transport. 

In addition, the manipulation of the light polarisation should allow controlling the one-by-one release of the NPLs. The two polarizations studied induced different dynamics and their effects are enhanced by the presence of the zero-order beam. For polarisation parallel to the trap the escape of the NPL is easier (faster) but their drift is not constant. For perpendicular polarization, the NPLs remain anchored to the trap center for a longer time (some even did not have time escape during our experiment) but their trajectories are smoother as they escape. 
Our FDTD simulations of Figure~\ref{fig:FDTD} revealed the enhancement of the electric intensity when a NPL is trapped in the optical line at the location of the zero-order beam. We expect the enhancement to increase by adding more NPLs. 
Simulations by coupling Electrodynamics-Langevin dynamics\cite{Sule2015} could shed light on how the collective dynamic of multiple NPLs reshape the optical field and optical trap, and more specifically how the symmetry of the assembly influences the NPL dynamic on each side of trap center. However preliminary results shown in the Supporting Information recapitulate the step-like trajectories observed in the experiment, the aspect ratio and irregular shape of the NPLs make the simulations very challenging. 

\textbf{Summary and conclusions}. We trapped and oriented anisotropic disk-like Au nanoplatelets in optical line traps. The unique electrodynamic interaction between the NPLs and their dynamics depend on their arrangement with respect to the polarization direction of the incident beam compared to the trap axis and the presence and abscence of an intense and tightly focused zero-order beam. Switching the sign of the optical phase gradient drives assembly or disassembly of the NPL arrays over distances of tens of microns. These findings are surprising and are not mere extensions of recent reports on the importance of the optical phase gradient force in controlling optical matter \cite{Roichman2008, Yan2015}. Since NPLs exhibit significantly slower transport compared to plasmonic nanospheres due to their specific mutipolar excitations, we are able to controllably clear contaminating nano-spheres from the line trap  revealing highly correlated motions at steady-state and during the transient assembly/disassembly processes. These correlations reflect near-field and optical binding type interactions of the NPLs and the optical trap, as supported by results of our FDTD simulations. In particular, the superposition of a focused Gaussian beam at the center of the line trap stabilizes the long NPL linear arrays and strongly influences the long-range spatial correlations of their motions during assembly and disassembly.

To our knowledge, there have been no similar observations of correlated drift reported for plasmonic nanospheres. If analogous correlated drift occurs in the driven dynamics of nanospheres its observation will require much higher frame rates or more viscous solution conditions to measure. In fact, we expect the collective drift phenomenon to be somewhat unique to NPLs (and perhaps other higly anisotropic nanoparticles\cite{Simpson2017} as the multipolar excitations and the interactions inherent therein create a more complex potential energy landscape. The separations of the NPLs summarized in Table~\ref{tbl:dis} are not values typical of optical binding ($d\approx \lambda/n$ with $\lambda=800$nm, $n=1.33$ leading to $d\approx=600$nm or $1200$nm), whereas nanospheres are dominantly separated by there well-defined distances or at near-field separations.\cite{Burns1989,Yan2014a,Yan2015} Our findings expand the manipulation of plasmonic nanoparticle interactions by exploiting temporal and spatial correlation between particles separated by some hundreds of nanometers to few microns. It opens opportunities for exploiting multipolar plasmon modes to sculpt light with nanoscale precision over tens of micrometers.

\begin{acknowledgement}
The authors acknowledge support from the Vannevar Bush Faculty Fellowship program sponsored by the Basic Research Office 
of the Assistant Secretary of Defense for Research and Engineering and funded by the Office of Naval Research through grant N00014-16-1-2502. Computer time at the Center for Nanoscale Materials, an Office of Science user facility, was supported by the U. S. Department of Energy, Office of Science, Office of Basic Energy Sciences, under Contract No. DE-AC02-06CH11357. We also thank the Research Computing center for an award that allowed conducting the FDTD simulations and modal analysis.We thank the University of Chicago NSF-MRSEC (DMR-0820054) for central facilities support. 
\end{acknowledgement}

\begin{suppinfo}

\end{suppinfo}
\bibliography{plateb}

\end{document}